# Analysis, Design and Simulation of a New System for Internet Multimedia Transmission Guarantee

O. Said, S. Bahgat, S. Ghoniemy, Y. Elawady

Computer Science Department
Faculty of Computers and Information Systems
Taif University, Taif, KSA.
dr_osaid@yahoo.com

*Abstract:* **QoS is a very important issue for multimedia communication systems. In this paper, a new system that reinstalls the relation between the QoS elements (RSVP, routing protocol, sender, and receiver) during the multimedia transmission is proposed, then an alternative path is created in case of original multimedia path failure. The suggested system considers the resulting problems that may be faced within and after the creation of rerouting path. Finally, the proposed system is simulated using OPNET 11.5 simulation package. Simulation results show that our proposed system outperforms the old one in terms of QoS parameters like packet loss and delay jitter.**

1. INTRODUCTION

The path that the multimedia streams is to follow should provide it with all required Quality of Services (QoS). Suppose that the determined multimedia path gives the multimedia streams all the needed services. In this situation, an urgent question arises. The question is what is the solution if, during the multimedia streams are transmitted in the path, that path is failed? This state may cause a loss in multimedia streams especially when are transported under the User Datagram Protocol (UDP). So, the solution is either to create an alternative path and change the multimedia streams away to flow in the new path or retransmit the failed multimedia streams. The second solution is so difficult (if not impossible) because the quantity of lost multimedia streams may be too huge to be retransmitted. So, the only available solution is to create another alternative path and complete the transmission process. To determine an alternative path, we face two open questions. The first question is: how a free path, that will transport the multimedia streams to the same destination, is created? The second question that may be put forward after the path creation is: can the created path provide the required QoS assigned for the failed one? From these queries and RSVP analysis, it's obvious that the elements of resource reservation and QoS are RSVP, routing protocol, sender, and receiver. Also, it's notable that the resource reservation process occurs before the multimedia transmission. At the beginning of the multimedia streams transmission, the relations between the QoS elements are disjoint. Hence, if a change occurs in the reserved path during the multimedia streams transmission operation, the previous stated problems may occur [1], [2].

In this paper, a new system for internet multimedia transmission guarantee is proposed and solves the old ones problem. This paper is organized as follows. In section 2, the related work that contains the RSVP analysis and DiffServ & MPLS evaluation is illustrated; in section 3, the problem definition is introduced; in section 4, our system is demonstrated; in section 5, detailed simulation and evaluation of our system are showed. Finally, the conclusion and the future work are illustrated.

2. RELATED WORK (RSVP, DIFFSERV, AND MPLS)

The three systems that are closely related to our work are RSVP, DiffServ, and MPLS. In this section, a brief analysis for RSVP is introduced. In addition, an evaluation of DiffServ & MPLS is demonstrated.

*A. RSVP operational model*

The RSVP resource-reservation process initiation begins when an RSVP daemon consults the local routing protocol(s) to obtain routes. A host sends Internet Group management Protocol (IGMP) messages to join a multicast group and RSVP messages to reserve resources along the delivery path(s) from that group. Each router that is capable of participating in resource reservation passes incoming data packets to a packet classifier and then queues them as necessary in a packet scheduler. The RSVP packet classifier determines the route and QoS class for each packet. The RSVP scheduler allocates resources for transmission on the particular data link layer medium used by each interface. If the data link layer medium has its own QoS management capability, the packet scheduler is responsible for negotiation with the data-link layer to obtain the QoS requested by RSVP. The scheduler itself allocates packet-transmission capacity on a QoS-passive medium, such as a leased line, and also can allocate other system resources, such as CPU time or buffers. A QoS request, typically originating in a receiver host application, is passed to the local RSVP implementation as an RSVP daemon. The RSVP protocol is then used to pass the request to all the nodes (routers and hosts) along the reverse data path(s) to the data source(s). At each node, the RSVP program applies a local decision procedure called admission control to determine whether it can supply the requested QoS. If admission control succeeds, the RSVP program sets the parameters of the packet classifier and scheduler to obtain the desired QoS. If admission control fails at any node, the RSVP program returns an error indication to the application that originated the request. However, it was found that unsurprisingly, the default best effort delivery of RSVP messages performs poorly in the face of network congestion. Also, the RSVP protocol is receiver oriented and it's in charge of setting up the required resource reservation. In some cases, to reallocate the bandwidth in a receiver oriented way could delay the required sender reservation adjustments [3], [4], see Fig. (1).

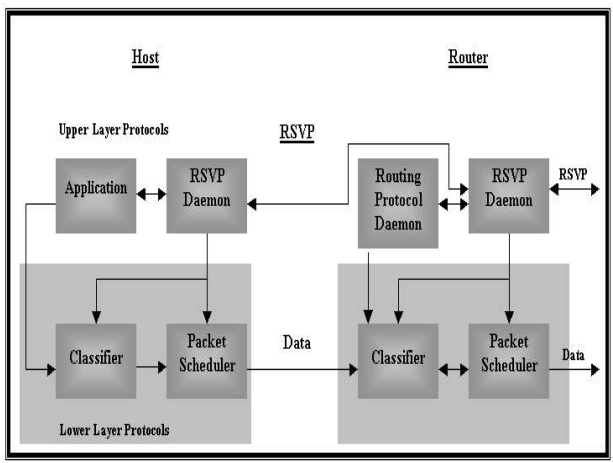

Fig. 1 The RSVP Operations

*B. DiffServ & MPLS*

MPLS simplifies the routing process used in IP networks, since in an MPLS domain, when a stream of data traverses a common path, a Label Switched Path can be established using MPLS signaling protocols. A packet will typically be assigned to a Forwarding Equivalence Class (FEC) only once, when it enters the network at the ingress edge Label Switch Router, where each packet is assigned a label to identify its FEC and is transmitted downstream. At each LSR along the LSP, only the label is used to forward the packet to the next hop.

In a Differentiated Service domain, all the IP packets crossing a link and requiring the same DiffServ behavior are said to constitute a behavior aggregate (BA). At the ingress node of the DiffServ domain, the packets are classified and marked with a DiffServ Code Point (DSCP), which corresponds to their Behavior Aggregate. At each transit node, the DSCP is used to select the Per-Hop Behavior (PHP) that determines the queue and scheduling treatment to use and, in some cases, drop probability for each packet [5], [6].

From the preceding discussion, one can see the similarities between MPLS and DiffServ: an MPLS LSP or FEC is similar to a DiffServ BA or PHB, and the MPLS label is similar to the DiffServ Code Point in some ways. The difference is that MPLS is about routing (switching) while DiffServ is rather about queuing, scheduling and dropping. Because of this, MPLS and DiffServ appear to be orthogonal, which means that they are not dependent on each other, they are both different ways of providing higher quality to services. Further, it also means that it is possible to have both architectures working at the same time in a single network, but it is also possible to have only one of them, or neither of them, depending on the choice of the network operator. However, they face several limitations:

1. No Provisioning methods
2. No Signaling as (RSVP).
3. Works per hop (i.e. what to do with non-DS hop in the middle?)
4. No per-flow guarantee.
5. No end user specification.
6. Large number of short flows works better with aggregate guarantee.
7. Works only on the IP layer
8. DiffServ is unidirectional – no receiver control.
9. Long multimedia flow and flows with high bandwidth need per flow guarantee.
10. Designed for static topology.

3. PROBLEM FORMULATION

The routing and resource reservation protocols must be capable to adapt a route change without failure. When new possible routes pop up between the sender and the receiver, the routing protocol may tend to move the traffic onto the new path. Unfortunately, there is a possibility that the new path can't provide the same QoS as the previous one. To avoid these situations, it has been suggested that the resource reservation protocol should be able to use a technique called the route pinning. This would deny the routing protocol the right to change such a route as long as it is viable. Route pinning is not as easy to implement as it sounds. With technologies such as Classless Inter-Domain Routing (CIDR) [7], [8], a pinned route can use as much memory from a router as a whole continent! Also, this problem may occur if a path station can't provide the multimedia streams with the same required QoS during a transmission operation. At this situation, the multimedia streams should search about an alternative path to complete the transmission process.

### 4. THE PROPOSED SYSTEM

From the problem definition and the RSVP analysis, it is obvious that the elements of the resource reservation and QoS are RSVP, routing protocol, sender, and receiver. Also, it is notable that the resource reservation process occurs before the multimedia transmission. At the beginning of the multimedia streams transmission (i.e. after the resources are reserved for the multimedia), the relations between the QoS elements are disjoint. So, if a change occurs in the reserved path during the multimedia streams transmission operation, the above stated problem may occur.

If the connections between the QoS elements are reinstalled during the multimedia streams transmission, then the QoS problems may be solved. The reinstallation process is accomplished by three additive components that are called the proposed system components.

*A. The proposed system components*

The proposed system comprises three additive components in addition to the old system components. The additive components are 1- Connector. 2- Analyzer. 3- Detector. In the following subsections, the definition and the functions of each additive component are demonstrated.

- Connector

This component is fired at the transmission starting and can be considered as a software class(s). The connector has more than one task for helping the system to accomplish its target. The main function of the connector is to reinstall the connections between QoS elements in a problem occurrence case, see algorithm 1 for more connector discussion.

- Analyzer

This component, located at the receiver, is considered also as a software class(s). The main function of the analyzer is to extract the failed station(s) and its alternative(s). Also, the analyzer connects to RSVP at the receiver site to extract a QoS request or a flow description of the new path. Also the analyzer uses some features of DiffServ and MPLS to acquire an alternative simple path with full QoS requirements. The DiffServ provides the system with simplest path and pushes the complexity to the network edges. The MPLS provides our system with next hop for each packet and to perform traffic conditioning on traffic streams flow in different domains (paths), see algorithm ٢ for more analyzer discussion.

- Detector

The detector and the connector are fired simultaneously. The detector precedes the connector in visiting the multimedia path's stations. The detector visits each path station to test the required QoS. If the detector notes a defect in the QoS at any station (i.e. the station can't provide the required QoS), then it sends to the connector an alarm message containing the station IP address and the failed required QoS, see algorithm 3for more detector discussion.

---

**Algorithm 1**

*1- While the number of multimedia packets < > Null*
  *2-1 Begin*
  *2-2 The multimedia starts the transmission operation*
  *2-3 The connector agent is fired with the starting of the transmission operation.*
  *2-4 For I = 1 To N.*
    *2-4-1 Begin*
    *2-4-2 The connector agent tests the stored detector flag value.*
    *2-4-3 If the flag value is changed to one.*
    *2-3-3-1 Go to the step number 3*
    *2-4-4 Else*
      *2-4-4-1 Complete the I For Loop*
    *2-4-5 End I For Loop.*
  *2-5 While ((SW-SC) * TR ) < > Null)*
    *2-5-1 Begin*
    *2-5-2 The connector extracts the nearest router address around the failed station.*
    *2-5-3 The connector sends a message to the router asking about alternative path (or station).*
    *2-5-4 The connector receives all available paths in a reply message sent by the router.*
    *2-5-5 The connector sends the router reply message to the analyzer asking about the new QoS request for the new path.*
    *2-5-6 For J = PFS To M*
      *2-5-6-1 Begin.*
      *2-5-6-2 The connector tests the QoS.*
        *2-5-6-2-1 If the QoS fails, the router returns to the step 2-5.*
        *2-5-6-2-2 Else, complete the J For Loop.*
      *2-5-6-3 End J For Loop.*
    *2-5-7 (SW-SC) * TR = ((SW-SC) * TR) –1(Unite time)*
    *2-5-8 End Inner while loop*
  *2-6 End outer while loop*
  *2-7 Stored flag value = 0.*
*2- End of the connector algorithm.*

**Algorithm 2**
*1- If the stored connector flag is changed to one*
    *2-1 The analyzer receives an old and a new paths from the connector.*
    *2-2 The analyzer compares between the two paths and separates the similar stations and the different ones.*
    *2-3 The analyzer keeps the similar stations in a table (called same) and keeps the different stations in another two tables (called Diff1 and Diff2).*
    *2-4 The analyzer constructs a mapping in relation to the QoS in the tables of different stations, see step 2.*
    *2-5 The analyzer cooperates with the RSVP to extract the QoS request of a new path.*
    *2-6 The analyzer capsulate the results in a message and sends it to the connector.*

*2- The analyzer handling and mapping operations*
    *2-1 For I = 1 to old[N].*
        *2-2-1 Begin*
        *2-2-2 If the old[I] = New[I]*
        *2-2-2-1 Begin.*
        *2-2-2-2 old[I] = Same[K]*
        *2-2-2-3 K=K+1*
        *2-2-2-4 End IF.*
    *2-2-3 Else*
        *2-2-3-1 Begin.*
        *2-2-3-2 old[I] = Diff1[H].*
        *2-2-3-3 old[I] = Diff1[H].*
        *2-2-3-4 H = H+1*
        *2-2-3-5 End Else.*
    *2-2-4 If H=K*
        *2-2-4-1 no changing in the old QoS request.*
    *2-2-5 For J = 1 to H*
        *2-2-5-1 Begin*
        *2-2-5-1 Diff2 [J] = Construct a QoS request.*
        *2-2-5-2 End J For Loop.*
    *2-2-6 End I For Loop*
*3- End of the analyzer Algorithm.*

**Algorithm 3**
*1- While the number of multimedia packets < > Null*
    *1-1 Begin*
    *1-2 If the QoS test value = 1*
        *1-2-1 Begin*
        *1-2-2 The detector multicast alarm message including the connector ID.*
        *1-2-3 The detector changes the test value to 0.*
        *1-2-4 The detector tests another succeed stations.*
        *1-2-5 End IF.*
    *1-3 End of the While Loop*
    *1-4 QoS test value = 0.*
*2- End of the detector algorithm.*

Note: the symbols description is found at appendix A.

Table 1: The Data Stored in each System Component

| Connector stored data | Analyzer stored data | Detector stored data |
|---|---|---|
| Connector ID | Connector ID | Detector ID |
| Address of each path station | Analyzer ID | Connector ID |
| Time of each visiting station | Connector address | Connector address |
| Analyzer ID | RSVP connections | QoS required from each path station |
| Analyzer Address | Similar table | Path structure |
| Stream ID | Different tables | The connector flag value (default value =0) |
| Detector flag value (default value =0) | The connector flag value (default value =0) | QoS test value (default value =0) |
| RSVP connections | | |

*B. System approach*

After the resource reservation processes have been done, the multimedia streams begin the flood across the predetermined path. The connector accompanies the multimedia streams at every station. When the connector receives an error message from the detector, the connector starts to install the connections between the QoS elements.

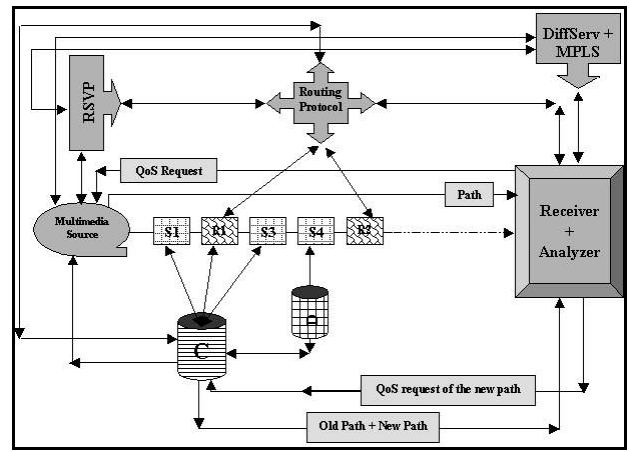

Fig. 2 Functional Diagram of the Proposed System

The connector extracts the address of the failed station and the nearest router. The connector constructs a message that will be sent to the routing protocol asking for an alternative path (or station). The routing protocol provides the connector with the available path(s) that compensates the old one. The connector constructs a message, containing the old and new paths, and sends it to the analyzer. The analyzer extracts the failed station(s) and its corresponding one(s) in the new path. The analyzer connects the RSVP to extract the QoS request. The analyzer constructs a message to be sent to the connector. The connector transforms the analyzer message to the sender informing it with the new selected path. Hence; the sender transmits the new multimedia streams using the new connector path see figures (2), (3).

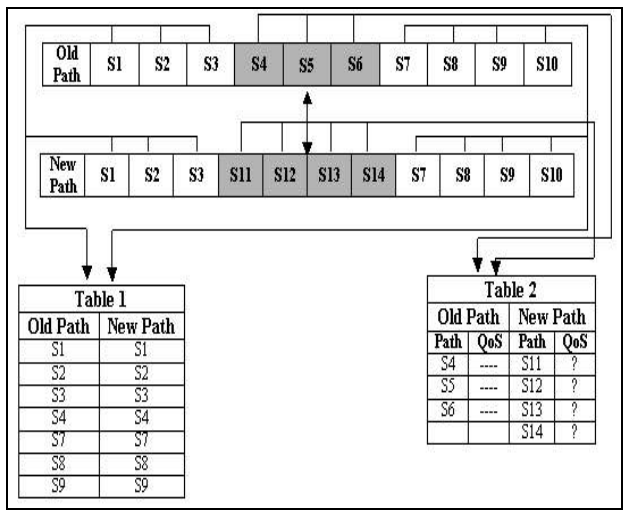

Fig. 3 Analyzer Operation

## C. System messages

To complete the connections between proposed system components, we have to demonstrate the structure of each used message. The proposed system contains five new messages that can be stated as follows.
1. From the connector to the sender.
2. Between the connector and the routing protocol (router) (Request and Reply).
3. Between the connector and the analyzer (Request and Reply).
4. Between the analyzer and RSVP at the receiver site (Request and Reply).
5. From the detector to the connector

- From the connector to the sender

This message joins the connector with the multimedia sender. This message is sent when the connector receives the QoS request from the analyzer. This message structure looks like the RSVP reservation request message but with the connector ID (This field is used in case of more than one connector in the proposed system).

- Between the connector and the routing protocol (Request and Reply).

This message joins the connector with the router or the routing protocol. This message is fired when the detector alarms the connector that a QoS failure is occurred at a station in the multimedia path. The connector needs this message to access the alternative path (or station) that replaces the failed path (or station). There are two types of this message, the request message and the reply message. The request message comprises the failed path and the reply message contains the alternative path. The request message has the following fields, 1) Message type, 2) Container ID, and 3) Old path. The reply message has the following fields, 1) Message type, 2) Connector ID, and 3) Alternative path(s).

- Between the connector and the analyzer (Request and Reply).

This message is used to communicate the connector and the analyzer. This message is fired when the connector needs a QoS request for the new path. The message has two types, the request message and the reply message. The request message contains a new path that is accessed from the router. The reply message contains the QoS request that is extracted after the analysis operation. The request message contains the following fields 1) Message type, 2) Container ID, and 3) Alternative path. The reply message contains the following fields 1) Message type, 2) Connector ID, and 3) QoS request.

- Between the analyzer and the RSVP at the receiver (Request and Reply).

This message is used to complete the dialog between the analyzer and the RSVP at the receiver site. The analyzer handles the old path and its alternative(s) to extract the failed station(s) and its corresponding station(s) in the new path. The analyzer needs it to construct a QoS request for the new path (s). This message has two types, the request message and the reply message. The request message contains the new path that was sent by the connector. The reply message contains the QoS request that is extracted by the RSVP. The request message contains the following fields, 1) Message type, 2) Analyzer ID, and 3) Alternative path. The reply message contains the following fields, 1) Message type, 2) Analyzer ID, and 3) Required QoS.

- From the detector to the connector

This message can be used to alarm the connector with a new event occurrence. If the detector finds a failure at a station in relation to QoS, then it sends this message to the connector asking to start its function for solving the problem. The message contains the following fields, 1) Message type, 2) Connector ID, 3) QoS request, and 4) Address of the failed station.

## D. Decreasing the number of system messages

It is notable that our system contains a number of messages that may cause a network overload. To make our system suitable for every network state, a new strategy to decrease a number of sent and received messages should be demonstrated. This strategy is built on the cumulative message idea. For the detector component, it's clear that its job is to test if each router (station) can provide the multimedia with required QoS or not. In case of network overload, the detector can capsulate its messages in one message. The capsulated message contains the addresses of the QoS failed stations that not visited by the multimedia streams in the transmission trip. For the analyzer component, it can use the same idea during the communication with the DiffServ and MPLS provided that the multimedia streams keep away from the analyzer transactions.

5. PERFORMANCE STUDY

In this section, the performance of the suggested multi-resource reservation system is studied. In our simulation the network simulator OPNET 11.5 [9] is used. A distributed reservation-enabled environment, with multiple distributed services deployed and multiple clients requesting these services is simulated. In particular, for runtime computation of end-to-end multi-resource reservation plans, the performance of the proposed system with the best effort communication system (old system) is compared. The key performance metrics in our simulations are: 1) End-to-end delay, 2) Packet loss, 3) Packet Loss in Case of Compound Services, 4) Re-Routing State, 5) Reservation Success Rate, 6) Utilization, and 7) Delay jitter. These parameters are evaluated for an increasing network load. Also, in our simulations, we compare between our system and the DiffServ && MPLS. In our simulation, Abhay Agnihotri study [10] is used to build the simulation environment.

*A. Simulation Setup*

The infrastructure of the simulation contains the following items:
1. 3 Ethernet routers to send and receive the incoming traffics and police it according to the QoS seniors specified in the proposed system, DiffServ, MPLS, and RSVP.
2. 15 video transmitters distributed on the router 1 and the router 2 as follows: 10 video transmitters are connected to router 1 and 5 are connected to router 2. The video workstations used to transmit 375 MPEG video packets per second, of size 1000 bytes. Each transmitter can send the multimedia packets only if it has a full required QoS like specified priority interactive, streaming, full bandwidth, specified delay jitter, and excellent effort.
3. 15 video receivers distributed on the router 2 and the router 3 as follows: 10 video receivers are connected to the router 2 and 5 are connected to the router 3.
4. The links between the workstations (video transmitters and receivers), are 1 Mbps. The links between the routers are 2 Mbps.
5. For internet simulation, the routers are connected via IP cloud.

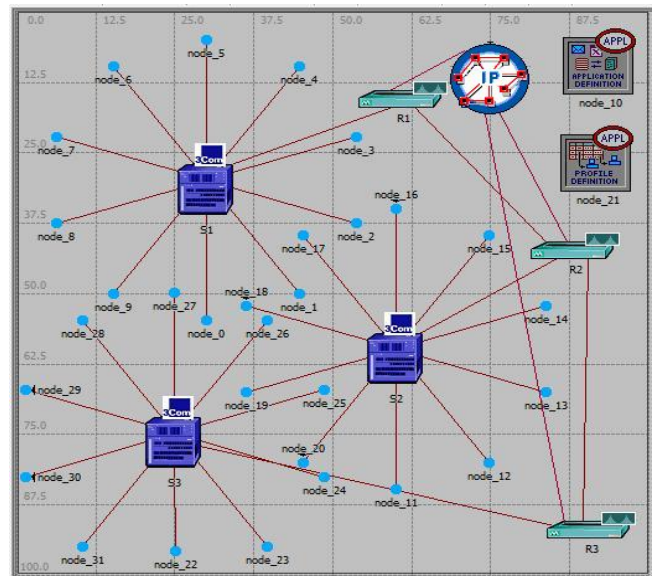

Fig. 4 Simulation Model Infrastructure

5.2 General Notes and Network Parameters

1. The data link rate and queue size for each queue scheme are fixed.
2. The multimedia traffics are considered MPEG with all characters.
3. The small queue size didn't affect the queue delay.
4. Inadequate data link rate causes more packet drops and too much consumed bandwidth.
5. Data link rate are fixed at 2 Mbps between the routers.
6. For FIFO queue scheme, the queue size was fixed at 400 packets.
7. The traffic pattern (continuous or discrete), the protocol (TCP or UDP), and application (prioritized or not) are considered input parameters.
8. The output parameters are calculated as regards the RSVP, DiffServ, MPLS, and our proposed technique.
9. It's supposed that the number of multimedia packets is increased with simulation time.
10. The simulation of the old system can be found at [11], [12].

*B. Simulation Results*

In our simulation, the parameters of multimedia and network QoS are scaled. The curves below contain a comparison between the old system (RSVP, DiffServ, and MPLS) and the new proposed system.

- End-to-End Delay

One of the key requirements of high speed packet switching networks is to reduce the end-to-end delay in order to satisfy real time delivery constraints and to achieve the necessary high nodal throughput for the transport of voice and video [13]. Figure 5 displays the end-to-end delay that may result from our computations, component messages and a buffer size. It's clear that our system computations didn't affect the delay time. This is because the computations are done during the multimedia transmission even a path failure is detected. Also, the old one uses the rerouting technique when finds a failure at any path station. The rerouting operations load the old system with more computations that will increase the time delay. In addition, our proposed system uses the cumulative message technique in case of network overflow.

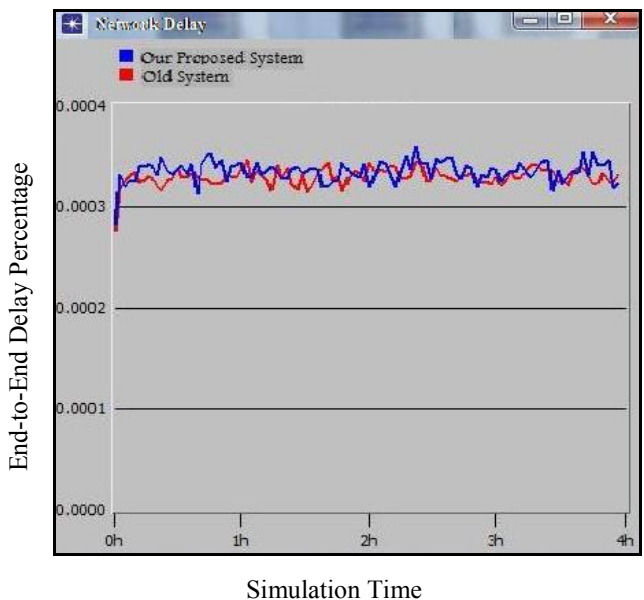

Simulation Time

Fig. 5 End-to-End Delay

- Packet Loss

This metric demonstrates the number of packet loss that occurred in the proposed system and the old system. The diagram found in figure 6 demonstrates the packet loss versus the time unit (it's supposed that the network load is increased with the time). It's obvious that the number of packet loss in our system is decreased compared to the old system. This decrease is justified by the following; the increasing in the network load means the increasing in the network hosts and this require services with different qualities. When the number of services and resources increases, the old system efficiency decreases hence; the number of packet loss increases. Unlike the old system, our system uses the detector, the connector, and the analyzer, to handle a failure that occurred in the old system before the multimedia packets affect and this promotes its efficiency. The number of packet loss is approximately equal especially before the middle of simulation time. The notable packet loss in our system comes from making the analyzer component inactive. The system fault tolerance will be discussed in the future work.

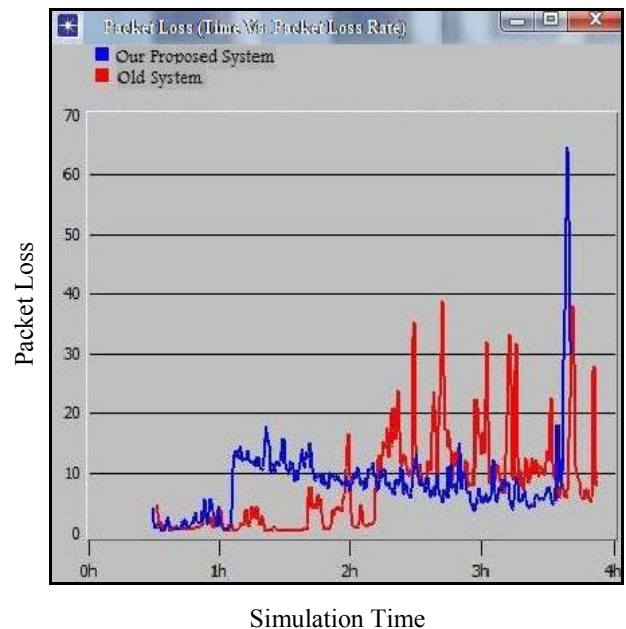

Simulation Time

Figure (6): Packet Loss

- Packet loss in case of compound services

This metric scales the efficiency of our system as regards the complete reservation of the resources that are required quality of the compound service. The compound service is a service that needs other one(s) to be reserved (dependant service). The curve in figure 7 shows the relation between the number of lost bits versus the generic times. It's notable that the efficiency of our system in compound service reservation is better than the old one. This indicates that the old system has a delay in dealing with the required compound services and this causes a loss of huge number of bits especially at the start of simulation time.

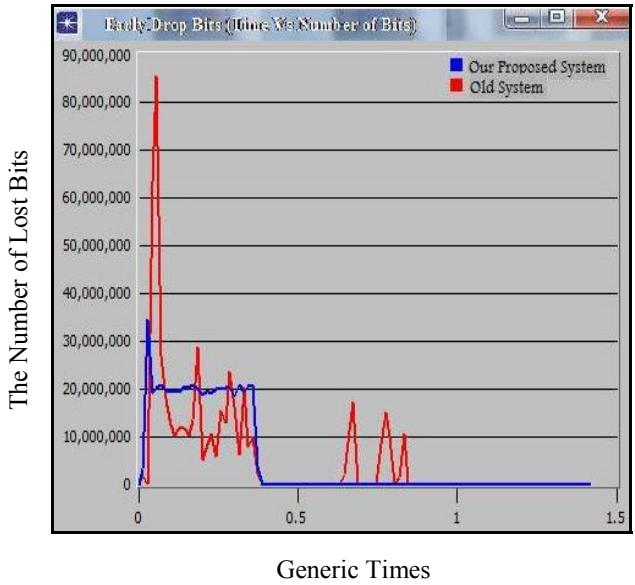

Generic Times

Fig. 7 Packet Loss in Case of Compound Services

- Re-Routing State

To meet high throughput requirements, paths are selected, resources reserved, and paths are recovered in case of failure. This metric should be scaled to make sure that our new system has an ability to find a new path when a failure occurred. This metric scales the rerouting state for our system and old one. The curve in figure 8 shows the relation between the number of recovered paths versus simulation time for new system and old one.

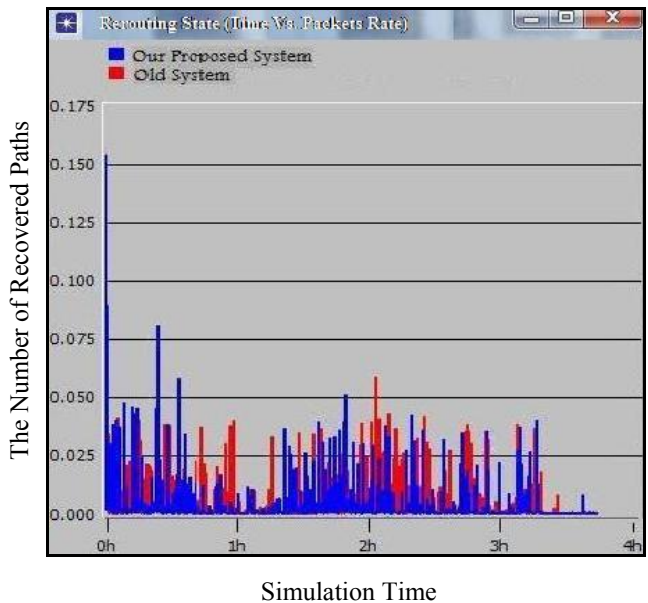

Simulation Time

Fig. 8 Re-Routing State

- Reservation Success Rate

This metric scales the efficiency of the proposed system as regard the resource reservation. The diagram in figure 9 shows the success reservation rates per time unit. It is observed that the success reservation rate in our system increases the success reservation rate in the old system. This increasing is due to efficiency of the detector in fault detection at any resource before it is used, in addition, efficiency of the connector in finding and handling the alternative solution. Also, the difference between the two systems is notable at the second hour of simulation time.

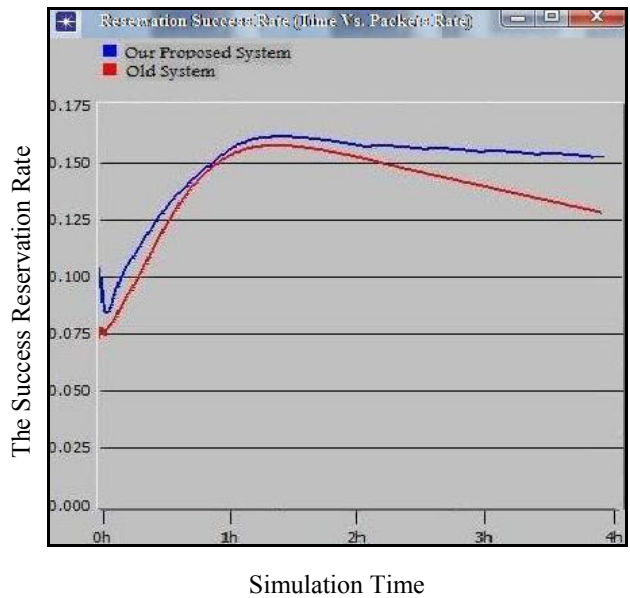

Simulation Time

Fig. 9 Reservation Success Rate

- Utilization

This metric scales the efficiency of our system additive components (the connector, the analyzer, and the detector). The efficiency of the connector is scaled by the number of successful connections in relation to the number of stations that cannot provide their QoS. The efficiency of the analyzer is scaled by the number of successful QoS requests extraction in relation to the number of its connections with the connector. The efficiency of the detector is scaled by the number of failed point's detection in relation to the number of failed points in the new system during the simulation time. For accuracy, all the components efficiency is scaled under different network loads. Figure 10 shows the average efficiency of three system components compared with the old system efficiency. The old system efficiency is calculated with a percentage of the services that are correctly reserved with the same required quality.

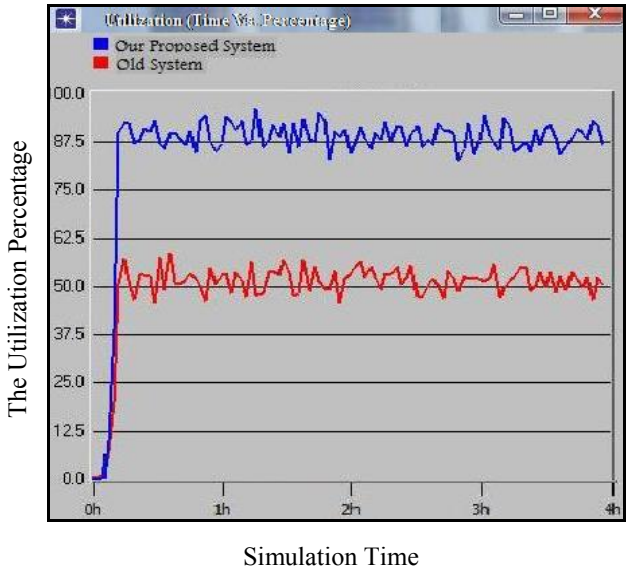

Fig. 10 Utilization (System Efficiency)

- Delay Jitter

This metric is introduced to make sure that the additive components didn't affect the multimedia packets delay jitter. The delay jitter as regards the multimedia streams is a very important QoS parameter. The plot in figure 11 describes the relation between the delay jitter and the first 1500 packets sent by the new system. In the new system's curve, it is obvious that the delay jitter is less than the old system's curve in the most simulation time. So, the additive components operate in harmony without affecting the delay jitter of the multimedia packets.

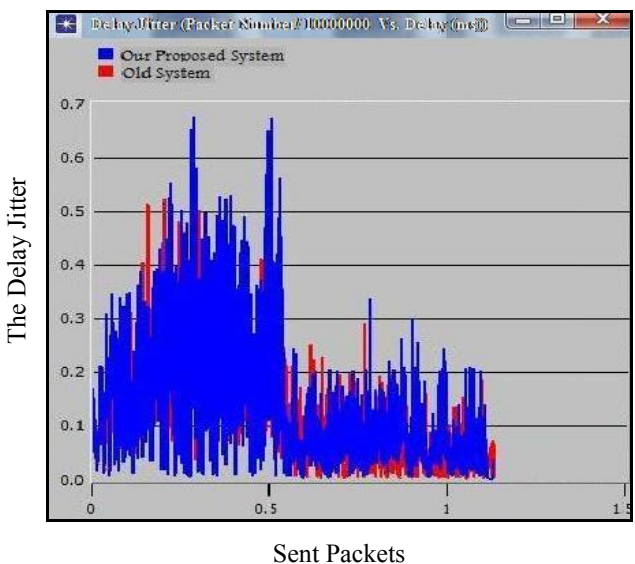

Fig. 11 Delay Jitter

## 6. CONCLUSION

In this paper, a brief analysis for the RSVP, the DiffServ, and the MPLS is demonstrated. Also, the QoS problems that may be occurred during the multimedia transmission are demonstrated. A new system to solve the QoS problem is introduced. The proposed system adds new three additive components, called connector, analyzer, and detector, over the old RSVP system to accomplish its target. A simulated environment is constructed and implemented to study the proposed system performance. A network simulator called OPNET 11.5 is used in the environment simulation construction. Finally, detailed comments are demonstrated to clarify the extracted simulation results. The test-bed experiments showed that our proposed system increases the efficiency of the old system with approximately 40 %

## 7. FUTURE WORK

To complete our system efficiency, the fault tolerance problem should be. What will be done if one of the system components fails? In our simulation, we faced this problem in packet loss diagram; hence we should find an alternative component (software or hardware) to replace the failed one and solve this problem. The suggested solution is to use a multi agent technology instead of one agent. Consequently, we simulate the multi agent QoS system and show the results. We will apply the proposed system with different types of multimedia data. This will make our system goes to the standardization. Hence, we can transform the proposed system to a new application layer protocol used for solving the multimedia QoS problems.

ACKNOWLEDGMENT

The authors would like to convey thanks to the Taif University for providing the financial means and facilities. This paper is extracted from the research project number 1/430/366 that is funded by the deanship of the scientific research at Taif University.

## Appendix A

**Assumptions:**

| Symbol | Description |
|---|---|
| SW | Detector visited station address. |
| SC | Connector visited station address. |
| TR | Time spent to reach any station. |
| TC | Connector visiting time. |
| I, J | Counters. |
| H, K | Two used variables. |
| PFS | Failed station position. |
| N | Number of stations in the old path. |
| M | Number of stations in the new path. |
| Old[ ] | Array used to keep the old path stations addresses. |
| New[ ] | Array used to keep the new path stations addresses. |
| Same[ ] | Array used to keep the similar stations found in the two paths. |
| Diff1[ ] | Array used to keep the different stations found in the old path. |
| Diff2[ ] | Array used to keep the different stations found in the new path. |

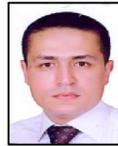
Dr. Omar Said is currently assistant professor in the Computer Science at Dept. of Computer Science, Taif University, Taif, KSA. He received Ph.D degree from Menoufia University, Egypt. He has published many papers at international journals and conferences. His research areas are Computer Networking, Internet Protocols, and Multimedia Communication

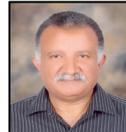
Prof. Sayed F. Bahgat is a Professor in the Department of Scientific computing, Faculty of Computer and Information Sciences, Ain Shams University, Cairo, Egypt. He received his Ph.D. from the Illinois Institute of Technology, Chicago, Illinois, U.S.A., in 1989. From 2003 to 2006, he was the head of Scientific Computing Department, Ain Shaams University, Cairo, Egypt. From 2006 to 2009 he was the head of Computer Science Department, Taif University, KSA. He is now a professor in the Computer Science Department, Taif University, KSA. Dr. Sayed F. Bahgat has written over 39 research articles and supervised over 19 M.Sc. and Ph. D. Theses. His research focuses in computer architecture and organization, computer vision and robotics.

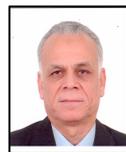
Prof. Said Ghoniemy is a Professor in the Department of Computer Systems, Faculty of Computer and Information Sciences, Ain Shams University, Cairo, Egypt. He received his Ph.D. from the Institute National Polytechnique du Toulouse, Toulouse, France, in 1982. From 1996 to 2005, he was the head of Computer Systems Department and director of the Information Technology Research and Consultancy Center (ITRCC), Ain Shaams University, Cairo, Egypt. From 2005 to 2007 he was the vice-dean for post-graduate affairs, Faculty of Computer and Information Sciences, Ain Shams University. He is now a professor in the Computer Engineering Department, Taif University, KSA. Dr. Ghoniemy has written over 60 research articles and supervised over 40 M.Sc. and Ph. D. Theses. His research focuses in computer architecture and organization, computer vision and robotics.

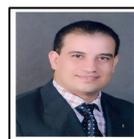
Eng. Yasser Elawady is a Lecturer in the Department of Computer Engineering, Faculty of Computers and Information Systems,Taif University,Taif, KSA. He received his M.Sc. from the Department of Computer Engineering, Faculty of Engineering, Mansoura university, Mansoura, Egypt, in 2003. His subject of interest includes *Multimedia Communication, Remote Access and Networking*.